# Structural Dependence of Chemical Durability in Modified Aluminoborate Glasses

Nerea Mascaraque[1], Kacper Januchta[1], Kristine F. Frederiksen[1], Randall E. Youngman[2], Mathieu Bauchy[3], Morten M. Smedskjaer[1,*]

[1]*Department of Chemistry and Bioscience, Aalborg University, Aalborg, Denmark*
[2]*Science and Technology Division, Corning Incorporated, Corning, USA*
[3]*Department of Civil and Environmental Engineering, University of California, Los Angeles, USA*
* Corresponding author. E-mail: mos@bio.aau.dk

**ABSTRACT**

Alkali and alkaline earth aluminoborate glasses feature high resistance to cracking under sharp contact loading compared to other oxide glasses. However, due to the high content of hygroscopic $B_2O_3$, it is expected that applications of these glasses could be hindered by poor chemical durability in aqueous solutions. Indeed, the compositional and structural dependence of their dissolution kinetics remains unexplored. In this work, we correlate the dissolution rates of aluminoborate glasses in acidic, neutral, and basic solutions with the structural changes induced by varying the aluminum-to-boron ratio. In detail, we investigate a total of seventeen magnesium, lithium, and sodium aluminoborate glasses with fixed modifier content of 25 mol%. We show that the structural changes induced by alumina depend on the network modifier. We also demonstrate a correlation between the chemical durability at various pH values and the structural changes in Mg-, Li- and Na- aluminoborate glasses. The substitution of alumina by boron oxide leads to a general decrease of chemical corrosion in neutral and acidic solutions. The lowest dissolution rate value is observed in Mg-aluminoborate glasses, as a consequence of the intermediate character of magnesium which can increase the network cross-linking. For basic solutions, the chemical durability is almost constant for the different amount of alumina in the three series, likely because $B_2O_3$ is susceptible to nucleophilic attack, which is favored in high-$OH^-$ solutions.

# I. INTRODUCTION

The commercial applications of oxide glasses can in many cases be limited by their chemical durability. Although borate-based glasses are good candidates for various applications, including solid-state batteries [1], special lasers [2], radiation dosimeters [3], seals [4], matrices for immobilization of high-level wastes [5] and biomaterials [6], they generally exhibit low chemical durability due to the high solubility of boron oxide in water [7,8]. There is thus a need for understanding the dissolution mechanisms and the compositional and structural dependence of chemical durability in acidic, neutral, and basic solutions. In oxide glasses [7,9,10], the main reactions, which can occur when the glasses are immersed in aqueous solutions, are (i) hydrolysis or the breaking of network former-oxygen bonds and, (ii) ion-exchange or diffusion of modifiers ($M^+$ and $M^{2+}$). Unlike network formers (such as $SiO_2$ or $B_2O_3$) that provide the network backbone, network modifiers (such as alkali or alkaline earth oxides) play a supporting role. This includes rupture of bridging bonds to create non-bridging oxygens or charge-balancers of tetrahedral network former units. The dissolution kinetics are correlated with the rigidity and connectivity of the glass network, as the penetration and attack of water molecules is less impeded in flexible compared to rigid glass networks [11,12].

In alkali borate glasses [13,14], two types of boron polyhedra, trigonal ($B^{III}$) and tetrahedral ($B^{IV}$) units, are present, and the alkali cations either serve to charge stabilize $B^{IV}$ or create non-bridging oxygens on $B^{III}$. Addition of alumina to borate glasses [15–18] changes the boron speciation, as tetrahedral aluminum also requires charge stabilization, either through alkali cations or five- and six-fold coordinated aluminum. For low modifier amounts (<30 mol %), the formation of non-bridging oxygen may be further suppressed by the presence of alumina [16]. The boron and aluminum speciations depend on the modifier type and content, and the Al/B ratio. Recently, Januchta et al. [19] demonstrated that the introduction of $Al_2O_3$ in sodium borate glasses can lead to improved hardness and, especially, crack resistance. Indeed, substitution of $Na_2O$ with $Li_2O$ led to the discovery of a $24Li_2O–21Al_2O_3–55B_2O_3$ glass that exhibits the highest crack resistance ever reported for any annealed oxide glass [20]. In our preliminary study [21], the effect of modifier substitution in glasses with fixed Al/B ratio on the structural and mechanical properties was investigated.

Magnesium aluminoborate glasses were found to exhibit the highest hardness and intermediate crack resistance. Magnesium cations can play the role of network-former [22] and, hence, be responsible for the increase of network cross-linking, as well as the increase of hardness. As such, crack resistant aluminoborate glasses could find interesting applications, but might be hindered by poor chemical durability in aqueous solutions due to the high content of hygroscopic $B_2O_3$.

The chemical resistance to water attack in aluminoborate glasses is expected to increase with increasing fraction of four-fold coordinated boron [8], higher degree of network cross-linking, and the substitution of B-O by Al-O bonds, as the former are more susceptible to nucleophilic attack [7,8]. However, the structural dependence of chemical durability in borate glasses containing alumina remains poorly understood. In this work, we perform nuclear magnetic resonance (NMR) spectroscopy measurements to determine the structural changes induced by the introduction of $Al_2O_3$ in borate glasses with the presence of mono- or divalent modifiers ($Li^+$, $Na^+$, or $Mg^{2+}$). The dissolution kinetics of these glasses are studied through the measurement of the weight loss and the leaching of magnesium, lithium and sodium modifier cations in aqueous solutions of varying pH. We show a correlation between the structural changes as a result of composition variation and the glass dissolution rates, providing an improved understanding of the glass dissolution mechanisms. This work is thus the first step towards the prediction and design of new aluminoborate-based glasses with specific chemical durability and mechanical properties.

## II. EXPERIMENTAL

### A. Sample preparation

We have investigated aluminoborate glasses with different modifiers and Al/B ratio. The three selected composition series (in mol%) were: (i) $25MgO \cdot xAl_2O_3 \cdot (100-x)B_2O_3$ with $x$=15, 20, 25, and 30; (ii) $25Li_2O \cdot xAl_2O_3 \cdot (100-x)B_2O_3$ with $x$=5, 10, 15, 20, 22.5, and 25; and (iii) $25Na_2O \cdot xAl_2O_3 \cdot (100-x)B_2O_3$ with $x$=5, 10, 15, 20, 25, 27.5, and 30. The glass series are referred to as MAB-$x$, LAB-$x$, and NAB-$x$, respectively (see Table 1). These seventeen glasses were prepared using the melt-quenching method by mixing batches of the following reagent grade materials: $MgCO_3$ (Acros Organics), $Li_2CO_3$ (Ph Eur. ≥99.5%, Merck), $Na_2CO_3$ (Ph Eur. ≥99.5%, Sigma-Aldrich), $H_3BO_3$ (Ph Eur. ≥99.5%, Sigma-Aldrich), and $Al_2O_3$ (≥99%, Sigma-Aldrich). The batches, with their stoichiometric amounts, were melted in a Pt-Rh crucible at 1050-1550 °C, depending on the composition, for 1 h in an electrically heated furnace. The melts were cast onto brass plates and the transparent glasses were annealed for 30 minutes at their respective glass transition temperature ($T_g$), which was determined by differential scanning calorimetry (DSC) using a STA 449C (Netzsch) with constant heating rate of 10 K/min under argon flow. The error in $T_g$ is within ±2 °C.

The chemical compositions of the annealed glasses were determined through inductively coupled plasma optical emission spectroscopy (ICP-OES). The results show an agreement between the nominal and measured oxide molar fractions (Table 1). We note that the glass-forming region (i.e., compositional region, for which glass formation is possible using standard melt-quenching technique) [23] in the investigated aluminoborate glasses depends on the type of modifier. In this work, we keep the modifier content constant (25 mol%) and vary the Al/B ratio, in order to limit the number of non-bridging oxygens (NBOs) in the network for optimal mechanical properties. With the applied melt-quenching method, we found that homogeneous glasses could be prepared in the compositional range of 15-30, 5-25, and 5-30 mol% $Al_2O_3$ for magnesium, lithium, and sodium, respectively. Therefore, the three series can only be directly compared in the range of alumina content between 15 and 25 mol%.

**B. Structural characterization**

$^{27}Al$ magic-angle spinning nuclear magnetic resonance (MAS NMR) experiments were conducted at a magnetic field of 16.4 T, corresponding to a resonance frequency of 182.34 MHz. Measurements were made using an Agilent DD2 spectrometer and Varian 3.2mm MAS NMR probe. Glasses were powdered and packed into 3.2mm outer diameter zirconia rotors, providing sample spinning of 22 kHz. Data were collected using a short radio frequency pulse width of 0.6 μs (π/12 tip angle) and recycle delays of 4 s to ensure uniform excitation and adequate relaxation of nuclei during signal averaging of 1000-2000 transients. $^{11}B$ MAS NMR data were collected at 11.7 T (160.34 MHz resonance frequency), with the exception of the NAB-20 glass, which was studied at 16.4 T (224.52 MHz resonance frequency) in an earlier study [24], in

conjunction with a Varian VNMRs spectrometer and 3.2mm MAS NMR probe. Powdered samples were contained in 3.2mm zirconia rotors and spun at rates of 20 kHz. Typically 400 scans were co-added, using 0.6 μs pulse widths (π/12 tip angle) and recycle delays of 4 s. All MAS NMR data were processed without extra line broadening (apodization) and plotted against the frequency standards of aqueous boric acid (19.6 ppm) and aqueous aluminum nitrate (0.0 ppm). Spectra were fit using DMFit [25] and quantitation of the different boron and aluminum sites was performed by integrating the fitted peak areas and applying a small correction for the overlapping satellite transition due to $B^{IV}$ sites [26].

$^{11}$B 3Q MAS NMR data were also collected at 11.7 T, using the standard three pulse, z-filtered experiment, where the hard 3π/2 and π/2 pulse widths were calibrated to be 2.8 and 1 μs, and following a short delay of 50 μs, a soft z-filter reading pulse of 20 μs was used. 24 to 48 transients were collected for each of 160 $t_1$ points, and then processed using Agilent VnmrJ software, which included proper shift referencing and shearing of the 2D datasets.

**C. Property characterization**

Density ($\rho$) of the annealed glasses was measured at room temperature using the Archimedes method. The weight of each glass was measured in ethanol using a balance measuring to ±0.1 mg. The measurement was repeated ten times for each glass. Based on the measured density, the atomic packing density ($C_g$) was calculated by assuming that the constituent atoms are spheres with known ionic radius [27]. That is, six-fold coordination for the alkali and alkali-earth cations, two-fold coordination for oxygen anions, and three- to six-fold coordination for B and Al atoms based on this work and our previous contributions [19,20]. For the modifier ions, we note that a detailed structural study would be necessary to determine the actual coordination number, which is not only six especially for $Mg^{2+}$. However, such determination is experimentally challenging and thus outside the scope of the present study. $C_g$ can be calculated as

$$C_g = \frac{\rho \Sigma f_i \left( 4/3\pi N_a (x r_A^3 + y r_B^3) \right)}{\Sigma f_i M_i}, \quad (1)$$

where for the *i*th constituent with chemical formula $A_xB_y$, $f_i$ is the molar fraction determined by ICP-OES and MAS NMR results (see Table S1 in Supplemental Material [28]), $r_A$ and $r_B$ are the ionic radii, $M_i$ is the molar mass, and $N_a$ is Avogadro's number.

For the dissolution rate measurements, samples of each aluminoborate composition were first cut to rectangular shapes and their faces were polished using 4000-grit SiC paper at the final step. Afterwards, they were cleaned with distilled water under ultrasonication for 5 min and then with acetone under ultrasonication for 5 min, and finally dried in an oven at 105 °C for 1 h. The weight and geometric surface area were measured. Aqueous solutions (50 mL) of 2 mM HCl, distilled water, and 0.5 M KOH were used to obtain pH values of 2, 7, and 14, respectively. For each experiment, the solid-to-liquid ratio was selected to be low enough to avoid any saturation effect. The bulk dissolution rate of each glass was determined by measuring the weight loss after immersion of the bulk samples in the three aqueous solutions at room temperature as a function of time (*t*). The experiment was done twice for each glass composition in each solution. The pH of the media was measured as a function of dissolution time using a pH meter (M220-X9541) and maintained constant throughout the experiments by adding basic or acid solutions. Note that the duration of the dissolution test was adjusted for each glass (from 32 h to 9 days) to achieve substantial values of weight loss. In addition to studying the bulk dissolution rates of the samples, the leaching of magnesium, lithium, and sodium modifier ions was also investigated after each immersion time. The concentrations of $Mg^{2+}$, $Li^+$ and $Na^+$ (in ppm) were determined using atomic absorption spectrometry (AAnalyst 100, Perkin Elmer).

## III. RESULTS

### A. Deconvolution of $^{27}$Al and $^{11}$B MAS NMR spectra

To determine how the addition of alumina affects the structure of the Mg, Li, and Na aluminoborate glasses, we have performed $^{27}$Al and $^{11}$B MAS NMR spectroscopy measurements. Due to the large number of NMR spectra (seventeen glasses) and the qualitative similarities, we here only show the $^{27}$Al and $^{11}$B MAS NMR spectra for the Mg-series (Fig. 1). The spectra for the remaining glasses (Li- and Na-series) are shown in Fig. S1 in the Supplemental Material [28]. The spectra are fitted using DMFit [25] and proper lineshapes for both quadrupolar nuclei. In the case of $^{27}$Al MAS NMR data for glasses, the CZSimple model, incorporating a Czjzek distribution of the quadrupolar interaction, provides accurate simulation of the highly asymmetric MAS NMR peaks (Fig. 1a) [29]. Fitting of $^{11}$B MAS NMR data is more straightforward in that 2$^{nd}$-order quadrupolar lineshapes are sufficient for all B$^{III}$ peaks, and due to the very small quadrupolar coupling constant, a combination of Gaussian and Lorentzian lineshapes accurately reproduces the B$^{IV}$ peaks (Fig. 1b).

The $^{27}$Al NMR spectra reveal three main peaks corresponding to Al$^{IV}$, Al$^{V}$, and Al$^{VI}$ at ~60, ~32, and ~5 ppm, respectively [30], while the $^{11}$B NMR spectra show two main peaks attributed to B$^{III}$ and B$^{IV}$ at ~17 and ~0 ppm, respectively [31]. However, the peak assigned to B$^{III}$ can be further deconvoluted into two distinct but overlapping peaks, which correspond to trigonal ring (B$^{III}_{ring}$) and non-ring (B$^{III}_{non-ring}$) units. The 3QMAS NMR data for $^{11}$B (see Fig. S2 in the Supplemental Material [28]) provides additional information for the two trigonal boron sites, including isotropic chemical shift, quadrupolar coupling and relative populations, which we used to validate the DMFit deconvolution of the $^{11}$B MAS NMR data. Table S1 in the Supplemental Material [28] summarizes the determined aluminum and boron speciation, along with their NMR-related parameters.

**B. Network former speciation**

Figs. 2a-c shows the composition dependence of the fraction of each aluminum and boron species in the three glass series, with the composition represented as [Al$_2$O$_3$]-[MO,M$_2$O] to enable comparison of the three series. Additionally, the composition dependence of boron and aluminum speciation of the three series combined in one plot is shown in Figs. S3a and S3b in the Supplemental Material [28]. The network former fraction is calculated from the chemical composition of the investigated glasses (Table 1) and the aluminum and boron speciation (see Table S1 in the Supplemental Materials [28]). Significant differences in the short-range structure are found as a function of modifier and alumina content. In the Mg-aluminoborate glasses, the fractions of both B$^{III}$ and B$^{IV}$ decrease and those of Al$^{IV}$, Al$^{V}$, and Al$^{VI}$ increase with increasing Al/B ratio (Fig. 2a). The decrease in B$^{III}$ fraction is higher than that for B$^{IV}$. However, the B$^{IV}$/(B$^{III}$+B$^{IV}$) ratio ($N^4$) is decreasing with increasing Al$_2$O$_3$ content, suggesting that B$^{IV}$ removal is more favorable upon the Al-for-B substitution. The fractions of Al$^{V}$ and Al$^{VI}$ are higher than that of Al$^{IV}$ in all four Mg-containing glasses. In the Li-series, the $B^{III}$ fraction decreases slightly more than that of $B^{IV}$ with increasing Al/B ratio, while all aluminum species exhibit similar rate of increase upon the addition of alumina (Fig. 2b). Comparison of the Mg- and Li-series reveals that the removal of the $B^{IV}$-species is more selective in the Li-series compared to the Mg-series. Moreover, the combined fraction of $Al^V$ and $Al^{VI}$ is lower than that of $Al^{IV}$ in the Li-series, which is also contrary to that found for the Mg-series (Fig. 2a). The Na-series exhibits qualitatively the same dependence of the network former fractions on Al/B ratio as the Li-series (Fig. 2c). However, the slight increase of $Al^V$ and $Al^{VI}$ species starts from NAB-20 glass ($[Al_2O_3]-[Na_2O]$ = -5 mol %) and LAB-15 ($[Al_2O_3]-[Li_2O]$ = -10 mol %) in Na- and Li- series, respectively. Moreover, there is a larger decrease of $B^{IV}$ and lower decrease of $B^{III}$ when $Al_2O_3$ is introduced in the Na-series compared to the Li-series. In conclusion, the Al-for-B substitution leads to a more efficient removal of $B^{IV}$ than $B^{III}$ units, but the propensity towards removing $B^{IV}$ is decreasing in the order Na>Li>Mg, i.e., with increasing field strength (ratio of charge to size) of the modifiers.

**C. Boron ring vs. non-ring units**

In sodium aluminoborate glasses [19], there is a preference for the modifier cations to charge-balance aluminum over boron tetrahedra, which is also the case for sodium boroaluminosilicate glasses [32]. However, the presence of five- and six-fold coordinated aluminum units suggests that more $B^{IV}$ species could be charge-balanced at the expense of $Al^{IV}$. In the borate network of the present glasses, the majority of the trigonal boron signal is assigned to $B^{III}_{ring}$ sites [15,33], while the remaining minor fraction is due to $B^{III}_{non-ring}$ sites [34,35] (see Fig. S4 in the Supplemental Material [28]). Note that we assume that both trigonal boron sites are only associated with bridging oxygens, in agreement with previous works [16,19,36]. In the Mg-series, most of the aluminum atoms are in five-fold coordination, suggesting that $Mg^{2+}$ is charge-balancing both aluminum and boron tetrahedra. Moreover, we have found that the abundance of $B^{III}_{non-ring}$ units is relatively higher in the Mg-aluminoborate glasses compared to the Na- and Li-aluminoborate glasses (see Fig. S4 in the Supplemental Material [28]). This suggests that the magnesium could cause breakage of $B^{III}_{ring}$ to $B^{III}_{non-ring}$ units by associating with the $B^{III}$ network through formation of $B^{III}$-O-Mg bonds, which is possible due to its partial network-former character with coordination number of four [21,31,32]. Thus, the pronounced difference in the structure of the Mg-series, compared to the Li- and Na-series for the same Al/B ratio, is due to the significantly higher field strength of $Mg^{2+}$ (0.45 $Å^{-2}$) compared to $Li^+$ (0.23 $Å^{-2}$) and $Na^+$ (0.17 $Å^{-2}$) and the dual nature of $Mg^{2+}$ in oxide glasses.

**D. Glass transition temperature**

The composition dependence of glass transition temperature ($T_g$), density ($\rho$), and atomic packing density ($C_g$) for the three glass series is illustrated in Fig. 3. The data is also summarized in Table 1. We observe an overall increase of $T_g$ (Fig. 3a) and $\rho$ (Fig. 3b) with [$Al_2O_3$] in the magnesium aluminoborate glasses. On the other hand, and as previously reported [19], the Li- and Na-glasses exhibit an initial decrease in $T_g$ with increasing amount of alumina, but $T_g$ then reaches a minimum around 20 and 25 mol% $Al_2O_3$, respectively. With additional $Al_2O_3$, $T_g$ increases for these glasses. In the $Al_2O_3$-rich region ([$Al_2O_3$]-[$M_2O$] ≥ 0), the $^{27}Al$ NMR results reveal the formation of $Al^V$ and $Al^{VI}$ species, which could be responsible for the higher degree of polymerization of the Al/B network, i.e, network becomes more rigid, and explains the sudden increase in $T_g$. In the Mg-series, the fraction of $Al^V$ and $Al^{VI}$ units are higher than that of $Al^{IV}$ and in addition, the magnesium can act as network former (e.g, four-fold coordination) in the studied compositional range, inducing a higher degree of network cross-linking and thus elevated $T_g$ values.

**E. Atomic packing density**

We next consider the trends in density values of the Li-and Na-series (Fig. 3b). In the range of compositions between 5 and 20 mol% $Al_2O_3$ ($B_2O_3$-rich glass), the atomic mass of sodium is higher than that of lithium, leading to a higher density for the Na-containing glasses. Nevertheless, the difference in density decreases with increasing alumina content ($Al_2O_3$-rich glass), which could be because the Li-glasses exhibit a more efficiently packed structure compared with the Na-glasses. The atomic packing density can illustrate the difference in free volume of the network, as $C_g$ represents the ratio between the minimum theoretical value of volume occupied by the constituent atoms and the molar volume of the glass. Indeed, Fig. 3c shows an initial decrease of $C_g$ with increasing alumina from 20 and 25 mol% $Al_2O_3$ in the Li- and Na-glasses. The packing density is associated to the average of coordination number, i.e, the change in five- or six- coordinated Al-species directly affects the $C_g$ values. In the $Al_2O_3$-rich regime of Li- and Na- series, there is an increase of $Al^{V,VI}$ species and the slope of $C_g$ decreases (a plateau of $C_g$ is observed, see Fig. 3c). For the Mg-containing compositions, the partial network-former character of magnesium could counteract the aluminum effect, and the $C_g$ values are approximately independent of Al/B ratio in the studied regime.

**F. Dissolution kinetics: weight loss curves**

We tested the chemical durability of the magnesium, lithium, and sodium aluminoborate glasses in acidic (pH = 2), neutral (pH = 7), and basic solutions (pH = 14). The weight loss normalized by the initial surface area of the glasses as a function of immersion time is shown in Figs. S5-S21 in the Supplemental Material [28]. When glasses are exposed to aqueous solutions, they corrode through various processes involving hydrolysis and hydration, which exhibit linear time dependence, and ion-exchange or diffusion, which exhibit square root time dependence [9]. The obtained weight loss curves for all seventeen glasses show a linear dependence on the immersion time, with initial non-linearity at short time, suggesting that surface hydrolysis is rate-limiting. Over the measured dissolution time, we observe that each solution remains far from saturation and no plateau in the weight loss vs. time plot is observed (see Figs. S5-S21 in the Supplemental Material [28]), indicating that the kinetics of the dissolution is only controlled by the glass structure, that is, with no retroaction effect from the solution.

**G. Dissolution kinetics: pH and composition dependence**

Fig. 4 shows the pH and composition dependence of the logarithmic dissolution rate $D_r$ for all the studied alkali and alkaline-earth aluminoborate glasses. $D_r$ is determined from the slope of linear fits to the weight loss curves in Figs. S5-S21 in the Supplemental Material [28]. The dissolution mechanism of borate-based glasses can be understood through the reactions of ion-exchange (Eqs. 2 and 4) and hydrolysis (Eqs. 3 and 5). In neutral solution, there is a mix of $H^+/H_3O^+$ and $OH^-$ ions, i.e., protonation and hydrolysis are strongly coupled with the possible reactions:

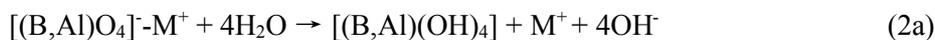

$[(B,Al)O_4]^- - M^+ + 4H_2O \rightarrow [(B,Al)(OH)_4] + M^+ + 4OH^-$ (2a)

$[(B,Al)O_4]^-_2\text{-}M^{2+} + 8H_2O \rightarrow [(B,Al)(OH)_4]_2 + M^{2+} + 8OH^-$  (2b)

$(B,Al)\text{-}O\text{-}B + H_2O \rightarrow 2\,((B,Al)\text{-}OH)$  (3)

Here, M represent a modifier atom. In acidic solutions, reaction (2) is favored as the ion-exchange step is a process of protonation of borate chains and there is higher concentration of $H^+$ present:

$[(B,Al)O_4]^-\text{-}M^+ + 4H^+ \rightarrow [(B,Al)(OH)_4] + M^+$  (4a)

$[(B,Al)O_4]^-_2\text{-}M^{2+} + 8H^+ \rightarrow [(B,Al)(OH)_4]_2 + M^{2+}$  (4b)

In basic solutions, reaction (3) is preferred because (B,Al)-O-B bonds are susceptible to nucleophilic attack of $OH^-$ (hydrolysis), which produces the following degradation:

$(B,Al)\text{-}O\text{-}B + OH^- \rightarrow (B,Al)\text{-}OH + B\text{-}O^-$  (5)

For all three glass series, we observe a decrease of dissolution rate ($D_r$) in acidic and neutral solutions with increasing $Al_2O_3$ content (Fig. 4). However, in basic media, the chemical durability (i.e., inverse of $D_r$) is approximately independent of Al/B ratio for the Mg-series (Fig. 4a) and slightly decreasing for Li- (Fig. 4b) and Na-series (Fig. 4c). The protonation process (Eqs. 2 and 4) is an ion-exchange reaction between the network modifying alkali or alkaline earth cation ($M^+$ or $M^{2+}$, respectively) and the protons ($H^+$), which is preferred in acidic media (reaction 4). The resistance to nucleophilic attack (Eqs. 3 and 5), i.e., breakage of B-O-B or Al-O-B bonds, depends on the rigidity of the borate network.

The Mg-glasses generally exhibit the lowest $D_r$ values for each pH in comparison to the corresponding Li- and Na-glasses (Fig. 4). For the Li- and Na-glasses with 5 to 15 mol% $Al_2O_3$ ($[Al_2O_3]-[M_2O] < -10$ mol%), similar $D_r$ values in all three solutions are observed (Figs. 5b-c). In this composition range, $B_2O_3$ is the major network-former, and the dissolution mechanism could be controlled by the hydrolysis process or the breakage of B-O-B bonds, as a consequence of the high solubility of $B_2O_3$ in water [7,8], independent of the solution pH. For the glasses with alumina content above 15 mol% ($[Al_2O_3]-[MO,M_2O] > -10$ mol%), the same pH and composition dependence of log $D_r$ is observed. That is, $D_r$ (pH=14) > $D_r$ (pH=2) > $D_r$ (pH=7), as well as a decrease of $D_r$ with increasing alumina content in all three glass series. The independence of $D_r$ on the Al/B ratio for pH=14 could be because the nucleophilic attack (Eq. 5) is favored at high pH. The structural dependence of chemical durability as a function of pH and glass composition is discussed in more details below.

**H. Dissolution kinetics: modifier leaching**

We have also studied the leaching of magnesium, lithium, and sodium modifier ions after each immersion time. This gives rise to a large dataset, so here we only show the time dependence of the concentrations of leached $Mg^{2+}$, $Li^+$, and $Na^+$ ions for the three glasses with 20 mol% $Al_2O_3$ in acidic, neutral, and basic solutions (Fig. 5). The leaching data for the remaining fourteen glass compositions are shown in Figs. S22-S24 in the Supplemental Material [28]. In addition to showing the measured modifier concentrations, we have also included the calculated modifier concentrations based on the weight loss measurements by assuming congruent dissolution. As shown in Fig. 5, we observe a linear increase in the concentration of dissolved $Mg^{2+}$, $Li^+$, and $Na^+$ ions with time for pH values of 2 and 7, with the measured values close to the calculated ones, i.e., congruent leaching of the modifiers for samples with 20 mol% $Al_2O_3$. However, non-congruent leaching of $Mg^{2+}$, $Li^+$, and $Na^+$ is observed at pH=14, especially for the Mg- and Li-glasses. Note that the large difference in the predicted and measured concentration of $Mg^{2+}$ could indicate the formation of magnesium oxide; however, no evidence of precipitation is observed in the basic solutions.

Considering the other glass compositions, we observe the same congruent leaching of $Mg^{2+}$ in acidic and neutral solutions for the studied range of alumina content (15-30 mol%) (see Fig. 5a and Fig. S22 in the Supplemental Material [28]). In basic solutions, the measured concentration of $Mg^{2+}$ is always lower than that calculated, likely because the magnesium ions can partly act as a network former [37,38] and are thus less able to be extracted from the glass network. Indeed, it appears that the degree of non-congruency at pH=14 increases with increasing field strength ($Mg^{2+}>Li^{2+}>Na^+$), as illustrated in Fig. 5. The leaching of $Li^+$ is congruent independent of the immersion time and alumina content (5-25 mol%) at pH values of 2 and 7 (see Fig. 5b and Fig. S23 in the Supplemental Material [28]). Except for the LAB-5 glass in acid solution, which exhibits non-congruent $Li^+$ release, the congruent leaching of $Li^+$ could be because part of these ions are acting to charge-balance $Al^{IV}$, in comparison to pure borate glasses, where $Li^+$ cations are charge-balancing $B^{IV}$. In basic solutions, the hydrolysis process (Eq. 5) is favored, in particular the breakage of B-O-B bonds for low [$Al_2O_3$], and there is thus less glass degradation by protonation process. The leaching of $Na^+$ is congruent in all the glasses with alumina content in the range of 15-30 mol% in neutral and acid solutions (see Fig. 5c and Fig. S24 in the Supplemental Material [28]). For the Na-glasses with lower alumina content (<15 mol%), the $Na^+$ release is non-congruent, likely because part of the $Na^+$ ions are preferentially charge-balancing $Al^{IV}$, as also observed for the LAB-5 glass.

## IV. DISCUSSION

The dissolution behavior is related to various factors such as atomic packing density, the rigidity of the network, modifier field strength ($Mg^{2+}$, $Li^+$ and $Na^+$), the content of network formers ($B_2O_3$ and $Al_2O_3$), and, in particular, the network former speciation (fractions of $B^{III}$, $B^{IV}$, $Al^{IV}$, $Al^V$, and $Al^{VI}$). A more rigid network is found with increasing modifier field strength in borate [39] and aluminoborate glasses [24], as high-field strength modifier cations form stronger bonds with oxygen [23,40]. Here, in agreement with the expected relation between field strength and dissolution kinetics, we find the highest $D_r$ values for glasses with high-field strength magnesium modifiers (Fig. 4). In this work, we use the structural information obtained from NMR studies to understand the trends in chemical durability of the aluminum-boron varying glasses.

In the compositional regime up to 15 mol% $Al_2O_3$, boron is the major network species with only a minor fraction of aluminum in predominantly four-fold coordination. In borate glasses, the $Li^+$ and $Na^+$ cations are charge-balancing $B^{IV}$, but the introduction of $Al_2O_3$ causes a fraction of them to charge-balance $Al^{IV}$, leading to a lower concentration of $B^{IV}$ and hence lower steric impediment in the $B_2O_3$ environment. Therefore, it is easier for water to penetrate the borate network, which is susceptible to nucleophilic reaction (Eq. 3), independent of the solution pH. This explains why the dissolution rates of Li- and Na aluminoborate glasses for pH values 2, 7, and 14 are similar (Fig. 4), when the water corrosion is mainly controlled by hydrolysis reactions (Eqs. 3 and 5). Note that $D_r$ values for pH = 2 in the Li-series are slightly higher than those in neutral and basic solutions, which could be because a higher fraction of $Li^+$ ions are acting as charge-balancer for $B^{IV}$ in comparison with NAB-5 and NAB-10 glasses, thus favoring the protonation reaction (Eq. 4).

In the compositional regime between 15 and 25 mol% $Al_2O_3$, the chemical durability in neutral and acid solutions increases with the addition of alumina for all three modifiers. However, the magnitude of the decrease in $D_r$ with the increase in $Al_2O_3$ content from 15 to 25 mol% in both neutral and acidic media is highest for the Na-series. To better understand the relation between structure and chemical durability, we determine and represent in Fig. 6 the variation in $D_r$ between the glass with 15 mol% $Al_2O_3$ and that with 25 mol% for pH = 2 and 7 and its corresponding variation in $N_B^4$ ($\Delta N_B^4$). Here $N_B^4$ is the fraction of tetrahedral to total boron species calculated from the $^{11}B$ MAS NMR data. Note that we select $N_B^4$ because the $B_2O_3$ network features high solubility in water, i.e., the boron fraction is found to be the best indicator of the glass dissolution rate. Fig. 6 shows that a higher variation of $N_B^4$ is directly related to a higher variation in $D_r$ at neutral and acidic pH values. The Na-series exhibits the highest $N_B^4$ change, i.e, the highest variation in $D_r$. In the three series, the presence of high aluminum oxide content in the borate network leads to a higher degree of polymerization, which directly affects the corrosion by hydrolysis, i.e, the higher the rigidity, the lower is $D_r$. Hence, the alumina addition can have a higher influence in neutral solutions, where the hydrolysis and protonation reactions are strongly coupled (Eqs. 2 and 3). This could explain why the variation in $D_r$ is higher at pH = 7 in the three series. Besides, these results confirm the important impact of the network former speciation, especially that of boron, on chemical durability in this compositional range. In borate glasses [8], $D_r$ decreases with increasing fraction of $B^{IV}$ units, but in the present aluminoborate glasses, the opposite behavior is observed. This might be related to the number of rigid bond constraints [11,12] associated with tetrahedral boron in these glasses. That is, preliminary modeling work suggests that although $B^{IV}$ has more two-body bond stretching constraints than $B^{III}$, it does not have any rigid three-body bond bending constraints. Additional work is needed to clarify this.

In the compositional regime above 25 mol% $Al_2O_3$, the Al/B network is more rigid and the hydrolysis and protonation reactions are thus impeded. As expected, the highest degree of network cross-linking blocks the penetration of water in the aluminoborate glasses, leading to higher $D_r$ for pH values of 2 and 7 in comparison to lower alumina containing glasses. On the other hand, we have found a lack of composition dependence for high pH dissolution, mainly due to the hydrolysis reaction (Eq. 5), which is favored in solutions with a higher concentration of OH⁻, together with the high solubility of $B_2O_3$, independently of the modifier.

## V. CONCLUSIONS

The structural dependence of dissolution kinetics in magnesium, lithium, and sodium aluminoborate glasses have been investigated by $^{27}$Al and $^{11}$B MAS NMR and weight loss measurements in acid (pH = 2), neutral (pH = 7), and basic (pH = 14) aqueous solutions. Additionally, glass transition temperature, density, atomic packing density, and modifier leaching measurements have also been performed. The introduction of $Al_2O_3$ gives rise to a general enhancement of the chemical durability in pH solutions of 2 and 7, which is directly related to the higher degree of network cross-linking and the arrangement of modifier in the glass Al/B network. In basic solution, the dissolution rates remain almost constant, independent of alumina amount in the three series, proving that the $B_2O_3$ network is highly susceptible to nucleophilic OH⁻ attack. The highest dissolution rates are found in aluminoborate glasses with magnesium, which can partially act as a network former, increasing network rigidity. The degree of congruency depends on the alumina and modifier contents and the solution pH. For $Al_2O_3$ content between 15 and 25 mol% (where the three series are comparable), a direct relation between the variation of $D_r$ and its corresponding variation in $N_B^4$ is found, i.e, the highest $N_B^4$ change leads to the higher $D_r$ increase. Finally, a congruent leaching of magnesium, lithium, and sodium ions is observed at pH values of 2 and 7, while for pH=14, the degree of non-congruency increases with increasing field strength.

## ACKNOWLEDGEMENTS

a
This work was supported by VILLUM Fonden (Postdoctoral Block Fellowship Program). M.B. acknowledges support from the National Science Foundation (Grant No. 1562066). We thank A. Goel (Rutgers University) for providing some of the samples.

**FIGURE CAPTIONS**

**FIG. 1.** (a,c) $^{27}$Al and (b,d) $^{11}$B MAS NMR spectra for the aluminoborate glasses in the Mg-series. The spectral deconvolution are shown in (a,b) along with the impact of Al/B ratio on the spectra in (c,d).

**FIG. 2.** Composition dependence of the fractions of aluminum and boron units in the (a) Mg, (b) Li, and (c) Na glass series. The former fractions are calculated based on the deconvoluted MAS NMR spectra and represent the amounts of the different network-forming units relative to the combined molar contents of $Al_2O_3$ and $B_2O_3$. Errors in the fractions are on the size of the symbols.

**FIG. 3.** Composition dependence of (a) glass transition temperature $T_g$, (b) density $\rho$, and (c) atomic packing density $C_g$ of the three compositions series. The errors associated with $T_g$, $\rho$ and $C_g$ are smaller than ±2 °C, ±0.001 g cm$^{-3}$, and ±0.001, respectively. Data for Na-glasses are taken from Ref. [19].

**FIG. 4.** Composition and pH dependence of the logarithmic dissolution rate ($D_r$, in mg dm$^{-2}$ h$^{-1}$) for (a) Mg, (b) Li, and (c) Na glasses. Lines are drawn as guides for the eyes. The vertical dashed lines separate two compositional ranges. For [$Al_2O_3$]-[$M_2O$] < 10 mol %, only lithium and sodium aluminoborate glasses could be prepared.

**FIG. 5.** Concentration of leached (a) Mg$^{2+}$ ions in MAB-20, (b) Li$^+$ ions in LAB-20, and (c) Na$^+$ ions in NAB-20 as a function of immersion time at pH = 2, 7 and 14. Calculated concentrations of Mg$^{2+}$, Li$^+$, and Na$^+$ are also shown based on the measured weight loss and by assuming congruent dissolution. Lines are drawn as guides for the eyes.

**FIG. 6.** Alumina-induced change in the dissolution rate ($\Delta\log D_r$) as a function of the alumina-induced change in tetrahedral boron site ($\Delta N_B^4$) for [$Al_2O_3$] = 15-25 mol % at pH=2 and pH=7. $\Delta\log D_r$ and $\Delta N_B^4$ are calculated from the variation between the values of log $D_r$ and $N_B^4$ for the glass with 15 mol% $Al_2O_3$ and those of log $D_r$ and $N_B^4$ for the glass with 25 mol% $Al_2O_3$, respectively. Lines are drawn as guides for the eyes.

**TABLE CAPTIONS**

**TABLE 1.** Analyzed chemical compositions (mol%), glass transition temperature ($T_g$), density ($\rho$), and atomic packing density ($C_g$) of the alkali and alkaline-earth aluminoborate glasses. The errors in $T_g$, $\rho$, and $C_g$ do not exceed ±2 °C, ±0.001 g cm$^{-3}$, and ±0.001, respectively.

**Figure 1**

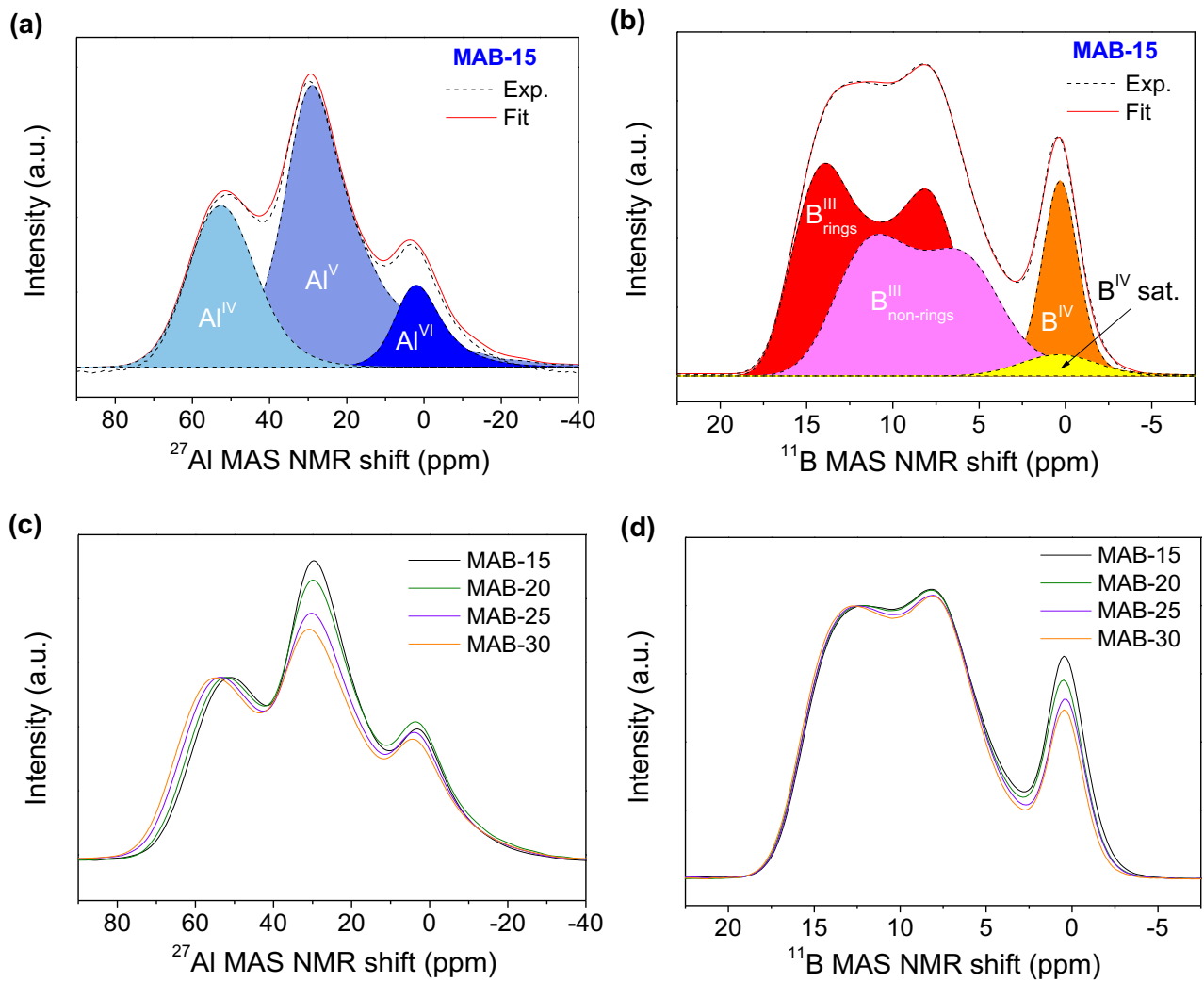

**Figure 2**

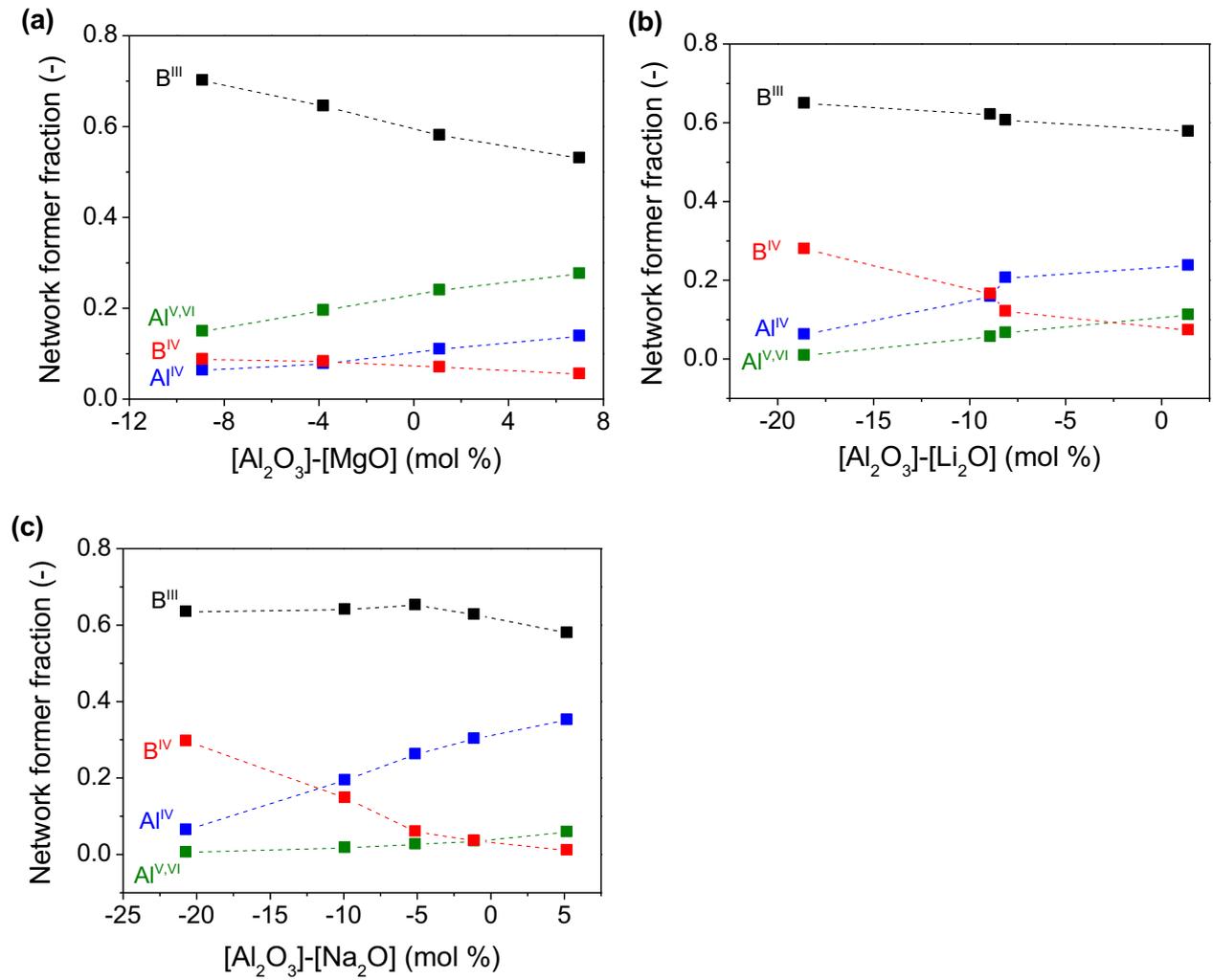

**Figure 3**

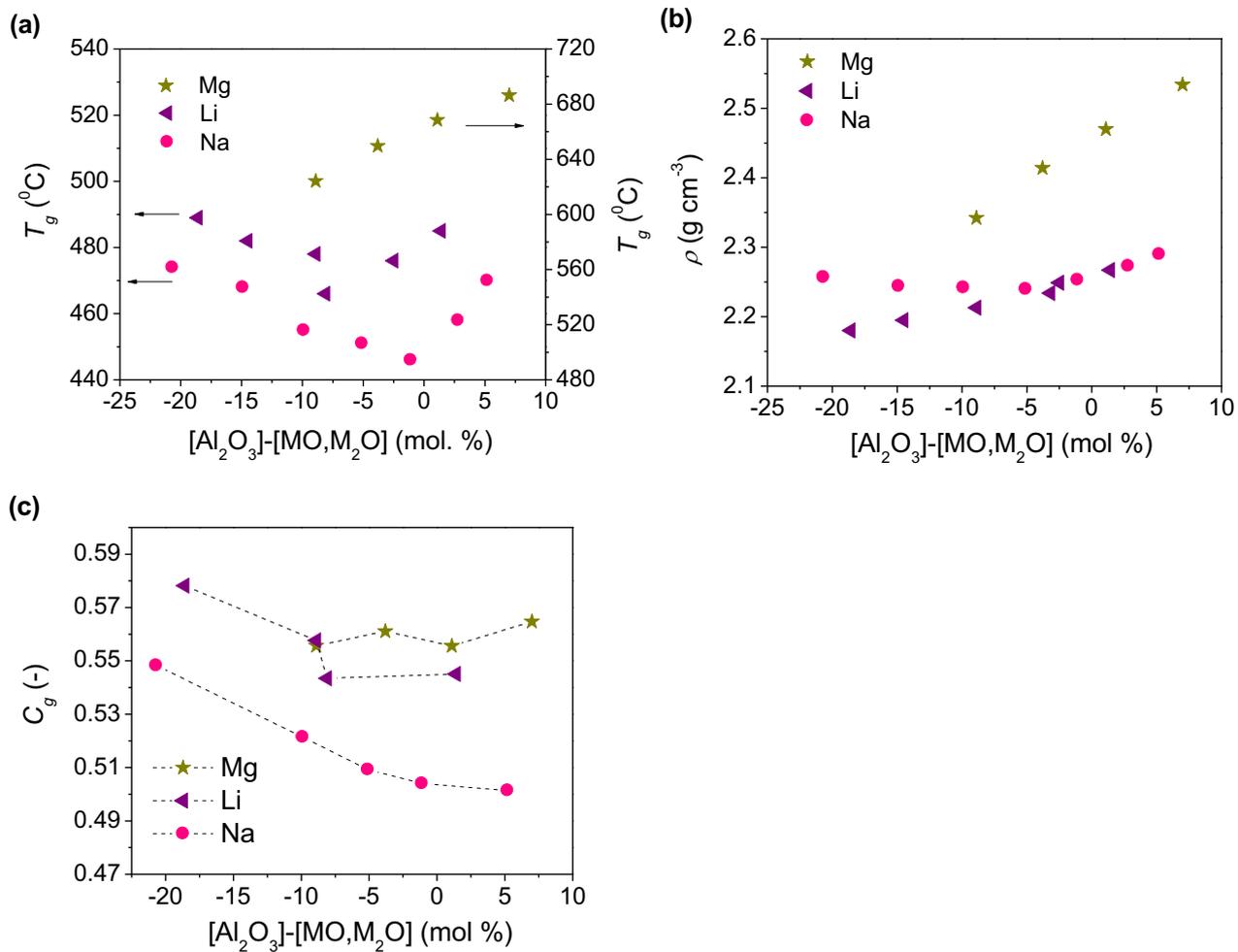

**Figure 4**

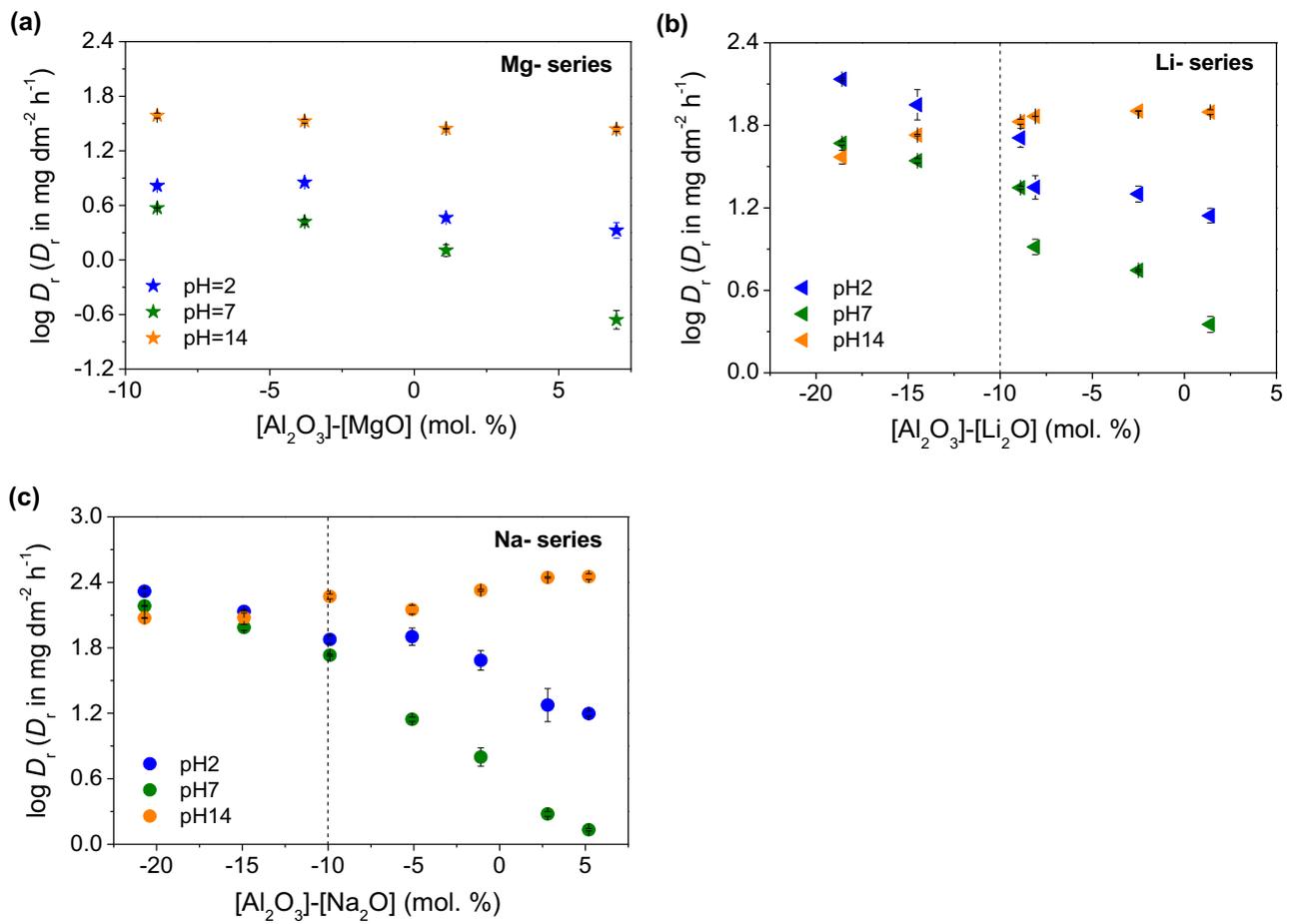

**Figure 5**

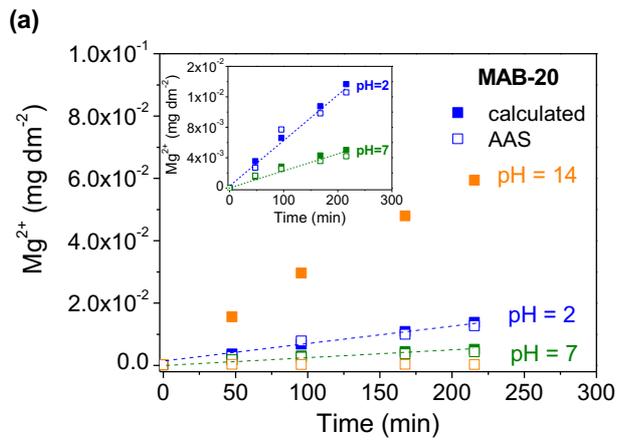
(a)

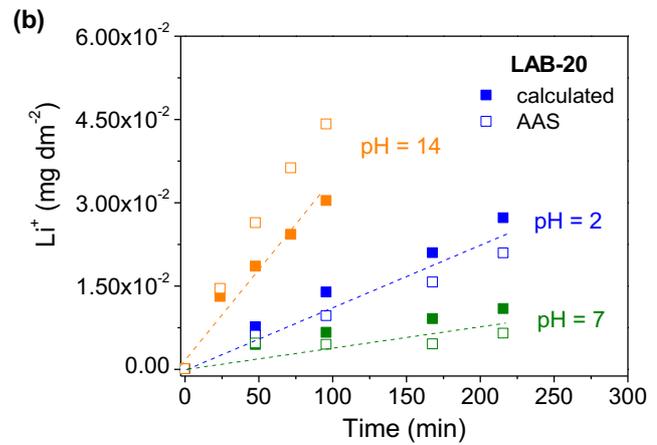
(b)

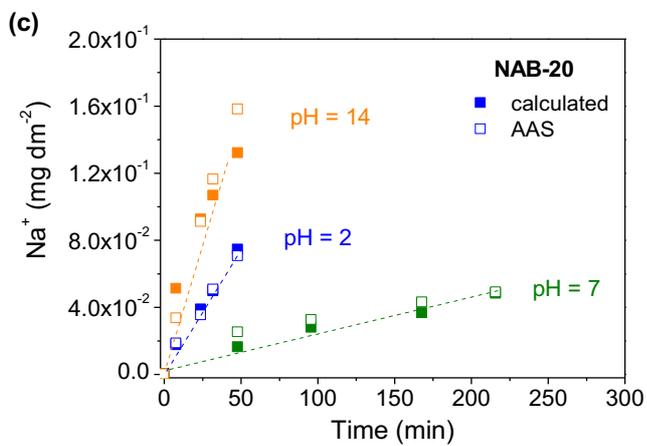
(c)

**Figure 6**

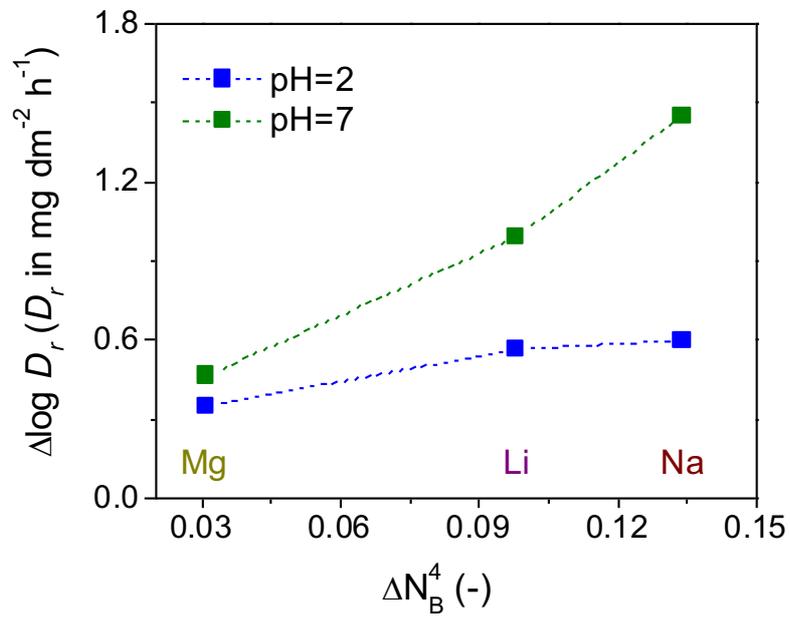

**TABLE 1**

| ID | Chemical composition (mol %) | | | | | $T_g$ (°C) | $\rho$ (g/cm³) | $C_g$ (-) |
|---|---|---|---|---|---|---|---|---|
| | MgO | Li$_2$O | Na$_2$O | Al$_2$O$_3$ | B$_2$O$_3$ | | | |
| MAB-15 | 24.2 | - | - | 15.3 | 60.5 | 624 | 2.342 | 0.553 |
| MAB-20 | 23.9 | - | - | 20.1 | 56.0 | 650 | 2.414 | 0.558 |
| MAB-25 | 24.7 | - | - | 25.8 | 49.5 | 668 | 2.470 | 0.543 |
| MAB-30 | 23.9 | - | - | 31.0 | 45.1 | 686 | 2.534 | 0.545 |
| LAB-5 | - | 23.9 | - | 5.3 | 70.9 | 489 | 2.180 | 0.578 |
| LAB-10 | - | 25.0 | - | 10.5 | 64.6 | 482 | 2.195 | - |
| LAB-15 | - | 24.7 | - | 15.8 | 59.5 | 478 | 2.213 | 0.558 |
| LAB-20 | - | 27.4 | - | 19.3 | 53.2 | 466 | 2.237 | 0.543 |
| LAB-22.5 | - | 24.7 | - | 22.2 | 53.1 | 476 | 2.249 | - |
| LAB-25 | - | 24.5 | - | 25.9 | 49.7 | 485 | 2.267 | 0.545 |
| NAB-5 | - | - | 25.7 | 5.0 | 69.3 | 474[a] | 2.257[a] | 0.548 |
| NAB-10 | - | - | 25.2 | 10.3 | 64.5 | 468[a] | 2.244[a] | - |
| NAB-15 | - | - | 25.2 | 15.3 | 59.5 | 455[a] | 2.242[a] | 0.521 |
| NAB-20 | - | - | 25.5 | 20.4 | 54.1 | 451[a] | 2.240[a] | 0.510 |
| NAB-25 | - | - | 25.5 | 24.4 | 50.1 | 446[a] | 2.253[a] | 0.504 |
| NAB-27.5 | - | - | 25.2 | 28.0 | 46.7 | 458[a] | 2.273[a] | - |
| NAB-30 | - | - | 25.0 | 30.2 | 44.8 | 470[a] | 2.290[a] | 0.501 |

[a] Data are taken from Ref. [19]